\documentclass[showpacs,preprintnumbers]{revtex4}
\usepackage{graphicx}
\usepackage{dcolumn}
\usepackage{bm}

\begin{document}

\title{General Relativity at an interface}
\author{Juan G. Diaz Ochoa}
\email{diazochoa@itp.uni-bremen.de}
\affiliation{\mbox{Institute for physics,Fachbereich 1, Bremen University,
Otto Hahn Alle, D28334 Bremen, Germany}}
\begin{abstract}
In this work a simple toy model for a free interface between bulk phases in space and time is presented, derived from the balance equations for extensive thermodynamic variables of Meinhold-Heerlein. In this case the free interface represents geodesics in the space-time, allowing the derivation of the Einstein's equations for gravitational fields. The effect of the balance equation is examined and a simple expression for cold dark matter is derived. The thermodynamically meaning of this model is also discussed. 
\end{abstract}

\maketitle

The observation of gravitational anomalies on visible matter has imposed a very difficult task due that the equations for gravitational fields require the introduction of additional non visible mass sources. The definition of dark matter as the source that originates these anomalies have stimulated a very interesting program for research the nature and composition of this dark matter. However, such deviations could also be a signal indicating that some important elements are lost in the fundamental equations of gravitational fields. 

Einstein's equations for gravitational fields describe the laws for the dependence of the metrical and gravitational field on space-time position, given a distribution of energy. It is interesting to point out that the relation between geometrodynamics and energy distribution is given by fundamental hydrostatic relations, describing the balance between energy and the geometry of a liquid in bulk. Many text books on general relativity derive these equations using hydrostatics; see for example A. Lichnerowitz \cite{Lichnerowicz}, R. Tolman \cite{Tolman} and K. Thorne et al. \cite{Gravitation}. The consideration of a bulk phase is essential in particular for the construction of cosmological models. However, the presence of interfaces separating different bulk phases in space and time is not excluded. Could the equations for gravitational fields be the effect of an interfacial dynamics more than a simple bulk phase? To answer this question is necessary to solve a very simple mathematical task by writing a kind of toy model for gravitational fields at an interface using a balance equation. 

Interfaces are very interesting objects which usually separates two inmiscible liquids. On the surface of an interface is usual to find fields depending on the perturbations of either of one or both of the phases. As a consequence, a model builded on this concept does not only depends on the particular coordinate system where an observation is done, but also depends on whether the observer approaches to the interface from one of both phases. Therefore, the information of the perturbations in the complementary phase can only be observed at the interface. This simple reasoning appears at the moment only an abstract formulation; but it could be a mechanism that also explains gravitational effects.  

The use of analogies between inviscid fluids and Riemanian manifolds is not new and have been explored with more or less intensity in the literature. For instance some theoretical works have examined how superfluid helium can be a model of space-times in general relativity, in particular the effects of superfluid A- superfluid B or 'impurity - superfluid' \cite{Konstantin}. These analogies are extended to models of superfluids at bulk in order to visualizes the vacuum as the ground state of condensed matter system  \cite{Chapline}, where ordinary matter can be visualized as analogous of exited states of this ground state; in such model the difference between the excited and ground states could explain the origin of dark matter. Randall and Sundrum presented a static solution to the 5D Einstein equations by means of the definition of the so called Brane World \cite{Gates}, where the space and time is flat on a 3D brane with positive tension given that the  bulk phase has a negative cosmological constant \cite{Stoica}. In this model the gravity is reproduced on the brane.

The present idea is grounded on the formulation of general surface conditions by considering a global balance equation in four dimensions; these conditions are commonly used in the calculation of the boundary of liquids with free interfaces, taking into account mass and energy transport across the interface, using a method developed by Meinhold-Heerlein \cite{Meinhold}. The equations of Meinhold-Heerlein was formulated in order to give a precise formulation of boundary conditions for interfaces of liquid-vapor systems in a very global form for a liquid that periodically evaporates and condensates at this interface. It has been applied for the calculation of the surface conditions of films of inviscid fluids, for example liquid helium \cite{Wehrum}, and complex liquids, for example films of polymers with temperatures about the glass transition temperature \cite{Diaz_01}. The interesting fact of this method is its geometric character because is based on fundamental concepts of differential geometry. 

The Meinhold-Heerlein equations are valid for extensible thermodynamic variables. The vectorial form of the equation for an extensible variable ${\psi }$, defined in both phases, can be calculated at the interface (without evaporation) as \cite{Meinhold} 
\begin{equation}
\overrightarrow{e\prime _{z}}\left( \overrightarrow{\Psi }_{1}-
\overrightarrow{u_{\Gamma }}\rho _{1}\psi _{1}\right) - \overrightarrow{e\prime _{z}}\left( \overrightarrow{\Psi }_{2}-\overrightarrow{u_{\Gamma }}\rho _{2}\psi _{2}\right) -\left( \frac{d\chi }{dt}-K{S}\right) + \tau  =0, \label{GEN_BALANCE}
\end{equation}
a balance equation at the interface with two fluxt densities $\overrightarrow{\psi}$ for the extensive variable, where $\overrightarrow{e\prime _{z}}$ is the vector normal to the interface, $\rho $ is the density of the corresponding phase, $\overrightarrow{u_{\Gamma }}$ is the velocity of the liquid at the interface, $\chi$ is the mass density of the interface, $S$ is the mean curvature, $K$ is a constant and $\tau$ is the source density of the extensive variable at the interface. These equations are in general valid for n-dimensional spaces. With this expressions it is possible to determine the equations at the surface for the balance of mass, linear momentum, energy and entropy of both phases.

An interface is usually assumed to be a bidimensional object between three dimensional bulk phases. In the model presented here the space is three dimensional surface between four dimensional bulk phases, such that the equations in general relativity are shifted from bulk into an interface. One must take into account that the linear momentum is a four vector that contains the energy on the time-coordinate. Therefore, the balance equation can be expressed in a tensorial form in the following way
\begin{equation}
(\Phi_{1}^{ij} - \Phi_{2}^{ij}) +u_{\Gamma }^{i}\left( \rho _{I}u_{I}^{j}-\rho _{I}u_{\Gamma }^{j}\right) -
u_{\Gamma }^{i}\left( \rho _{II}u_{II}^{j}-\rho _{II}u_{\Gamma}^{j}\right) - 
\left(g^{ij}\frac{d\chi}{dt}-KS^{ij}\right)+\tau^{ij} = 0,
\label{Meinhold}
\end{equation}
where $\Phi$ in this context is a tensor of tensions. Given that a four linear momentum is considered, then the following equation holds
\begin{equation}
\psi=1;
\end{equation}
furthermore, the substantial current density will be equivalent to the tensor of tensions. In a first instance a very simple model is presented with a liquid without internal interactions; therefore the tensor of tensions is equal to zero.
 
The energy tensor contains information of the energy and the tensions. The present description is for a balance of extensive variables at the interface. Therefore, the energy tensor in the present approximation does not contains information of interactions and is restricted only to the mass in the configuration. That means, the substantial current density will be equal to zero. The resulting balance equation for mass-energy is
\begin{equation}
u_{\Gamma }^{i}\left( \rho _{I}u_{I}^{j}-\rho _{I}u_{\Gamma }^{j}\right) - u_{\Gamma }^{i}\left( \rho _{II}u_{II}^{j}-\rho _{II}u_{\Gamma}^{j}\right) -
\left(g^{ij}\frac{d\chi}{dt}-KS^{ij}\right)+
\tau^{ij}=0.
\label{BALANCE_01}
\end{equation}
In the four dimensional case, $S^{ij}$ represents the curvature tensor defined by the equations of geometrodynamics. We assume in this equation that the system at bulk is at rest, i.e. $u_{I}^{j}=u_{II}^{j}=0$. Therefore, the interface is the single object in the system that is sensitive to any mechanical response. Defining
\begin{equation}
g^{ij} \frac{d\chi}{dt} + \tau^{ij}=g^{ij}\Lambda,
\end{equation}
the balance equations (\ref{BALANCE_01}) reduces to
\begin{equation}
u_{\Gamma }^{i}\rho _{II}u_{\Gamma }^{j}-u_{\Gamma }^{i}\rho _{I}u_{\Gamma
}^{j}+KS^{ij}+g^{ij}\Lambda =0.
\end{equation}
The energy tensor can be defined as (See Lichnerowicz \cite{Lichnerowicz} \& Tolman \cite{Tolman})
\begin{equation}
Q^{ij}=\left( u_{\Gamma }^{i}\rho _{II}u_{\Gamma }^{j}-u_{\Gamma
}^{i}\rho _{I}u_{\Gamma }^{j}\right) 
\end{equation}
where $K$ is a constant. The Einstein's equations for gravitational fields with sources can be easily recovered
\begin{equation}
Q^{ij}=KS^{ij}+g^{ij}\Lambda.
\label{Balance_02}
\end{equation}
If one consider, first that there is no surface source density, and second that the surface density remains constant, then the equation (\ref{Balance_02}) can be expressed as,
\begin{equation}
Q^{ij}=KS^{ij},
\label{Balance_03}
\end{equation}
which are the Einstein's equations without the cosmological constant. 

A first characteristic of the previous derivation is the possibility of introduce, in an elegant way, elements in the equation that has been introduced at the present in an ad-hoc way. A clear example is the cosmological constant. If the present toy model is consistent, then the cosmological constant can be defined as the whole density of the interface. Moreover the coupling to the variation in the surface density could imply that this apparent constant is in reality not a constant at all.

Here $Q^{ij}$ is given by the balance of energy at the interface. One is interested in the effect that this balance of energy can have in the equations for gravitational fields. In this case we approximate eq. (\ref{Balance_03}) to a Newtonian limit \begin{equation}
\frac{\partial \Gamma _{44}^{i}}{\partial q^{i}}=\frac{\kappa c^{2}}{2}(\rho
_{II}-\rho _{I})=\nabla ^{2}\phi,
\end{equation}
where $\phi$ is the gravitational potential. The balance of energy gives an important element in these equations. We assume that astronomical observations are done from a bulk phase, called 'reference bulk phase', close to the interface; this assumption implies that any gravitational field is the effect of the perturbation of the interface. Therefore, the observed geodesics are elements defined near the interface. The definition of an interface implies that the second bulk phase is hidden by this interface. Hence, there is a limitation of the information in the system. The space-time only allows the transmission of information at the interface and hides the transmission of the information in bulk (in a similar way as a common interface in a liquid-vapor phase can only reflect the light). The space-time can be considered as an interface that emerges from two phases with an effective free energy, which can be indirectly estimated (In other words, any detection is limited by the horizon defined by the interfacial state). 

The reference bulk-phase can be defined with a constant density, but variations in the second bulk phase can take place, particularly near the interface. In such a case if small variations in the density take place for example on the radial coordinate, they can be expanded in taylor series as 
\begin{equation}
\rho _{II}=\rho _{0II}+\frac{\partial \rho _{II}}{\partial q^{1}}\Delta \rho
_{II}+....;
\end{equation}
this expansion is restricted only to the first elements. Assuming that (See Navarro et al. \cite{Navarro})
\begin{equation}
\frac{\partial \rho _{II}}{\partial q^{1}}\delta \rho _{II}=\frac{\delta
_{c}\rho_{II} }{\left( \frac{q^{1}}{q_{s}^{1}}\right) \left( 1-\frac{q^{1}}{q_{s}^{1}}\right) ^{2}},
\end{equation}
where $\delta_{c}$ is a concentration parameter, $q_{s}^{1}$ is a scaled radius as a function of the mass of a halo of dark matter, and 
\begin{equation}
\rho _{0II}=\rho _{0I}, 
\end{equation}
then
\begin{equation}
\frac{\delta _{c}\rho_{II} }{\left( \frac{q^{1}}{q_{s}^{1}}\right) \left( 1-\frac{%
q^{1}}{q_{s}^{1}}\right) ^{2}}\approx \nabla ^{2}\phi.
\end{equation}
Therefore, from radial perturbations in the second phase at the interface is possible to obtain conventional equation for dark matter in newtonian approximation. A toy model using a concept of an interface for the whole space-time can explain actual limitations in observational astronomy. 

Given that in these equations the interface is the source of the gravitational field, then for a perfect balance of energy at the interface, 
\begin{equation}
S^{ij}=0,
\end{equation}
which is the classical equations for flat spaces \cite{Lichnerowicz}\cite{Tolman}.

 The curvature could also depends on the order parameters of each of the bulk phases. This control parameter can unfortunately not be controlled from the interface or from either of one bulk phases. If this dependence takes place, then some critical phenomena can happen at the space-time interface in a similar way as it take place in conventional condensed matter, implying a dependence of the gravitational potential on this order parameter. Therefore
\begin{equation}
\phi_{G} = \phi_{G}[{T_{G}}](r^{i}).
\end{equation}
Regarding recent astrophysical observations, the introduction of a single parameter avoids the problem of the definition of additional sources of energy in order to explain gravitational anomalies, preserving the structure of a simple balance equation. Furthermore, such a parameter could give a deep meaning of the interpretation of astronomical observations. The main problem in this context is to explain what exactly means a bulk phase. The answer of this problem could be found with the program proposed by Lee Smolin using the concept of quantum loops by explaining the microscopic nature of the bulk phases \cite{Rovelli}. Assuming the model presented here is plausible, gravitation could be, not a fundamental field, but an emergent phenomena.

I thank Johanes Schneider for very interesting discussions and Elena Ramirez for the critiques and revision of this manuscript. 


\end{document}